# PERIODIC ARTIFACT REDUCTION IN FOURIER TRANSFORMS OF FULL FIELD ATOMIC RESOLUTION IMAGES

Robert Hovden[1], Yi Jiang[2], Huolin L. Xin[3], Lena F. Kourkoutis[1,4]

[1] *School of Applied and Engineering Physics, Cornell University, Ithaca, NY 14853*

[2] *Department of Physics, Cornell University, Ithaca, NY 14853*

[3] *Center for Functional Nanomaterials, Brookhaven National Laboratory, Upton, New York 11973*

[4] *Kavli Institute at Cornell for Nanoscale Science, Ithaca, NY 14853*

**Corresponding Author**

Robert Hovden

271 Clark Hall, Cornell University, Ithaca, NY 14853

Tel: 607 255-0654

Fax: 607 255-7658

E-mail: rmh244@cornell.edu

## ABSTRACT

The discrete Fourier transform is among the most routine tools used in high-resolution scanning / transmission electron microscopy (S/TEM). However, when calculating a Fourier transform, periodic boundary conditions are imposed and sharp discontinuities between the edges of an image cause a cross patterned artifact along the reciprocal space axes. This artifact can interfere with the analysis of reciprocal lattice peaks of an atomic resolution image. Here we demonstrate that the recently developed Periodic Plus Smooth Decomposition technique provides a simple, efficient method for reliable removal of artifacts caused by edge discontinuities. In this method, edge artifacts are reduced by subtracting a smooth background that solves Poisson's equation with boundary conditions set by the image's edges. Unlike the traditional windowed Fourier transforms, Periodic Plus Smooth Decomposition maintains sharp reciprocal lattice peaks from the image's entire field of view.

## INTRODUCTION

Modern electron microscopists have the luxury of studying the atomic structure of materials directly through real space imaging. With resolving power that distinguishes atomic columns in a crystal, one can count the atoms across a grain boundary, identify the structure of interfaces, and locate individual defects and dopants (Wall et al., 1974; Cowley & Smith, 1987; Batson et al., 2002; Voyles et al., 2003; Muller, 2009). However, real-space analysis of crystals is not always sufficient and the discrete Fourier transform (DFT) of atomic resolution images is routinely used to mimic a diffraction pattern—with reciprocal lattice spots reflecting the symmetry and spacing of a specimen's crystal structure (Rust, 1974; Hashimoto et al., 1980).

In a diffraction experiment, electrons scattered through a crystal are recorded in the diffraction plane (Thomson & Reid, 1927). This diffractogram reveals the Fourier transform magnitude of the electron potential as $|F\{V(\mathbf{r})\}|^2$. On the other hand, a high-angle annular dark field scanning transmission electron micrograph (HAADF STEM) to a first approximation measures the real space atomic potentials, $|V(\mathbf{r})|^2$, (under an incoherent linear imaging model). Its Fourier transform, $F\{|V(\mathbf{r})|^2\}$, therefore, reflects the structure of a diffraction pattern but may permit additional peaks. Referring to the Fourier transform as a diffractogram can be misleading—or more insidiously, miseducating.

Moreover, taking the Fourier transform of a digital image introduces artifacts that can impede interpretability of the spectrum. The most noticeable artifacts present in the power spectrum of an electron micrograph are obtrusive cross pattern streaks running vertically and horizontally through the center of the 2D Fourier transform. These originate from the periodic boundary conditions imposed by the DFT. Imagine wrapping the edges of the image—top to bottom; left to right (as illustrated in Figure 1). The mismatch in intensity where the boundaries meet are sharp and discontinuous; spanning low to high frequencies in reciprocal space along the two principle axes directions (Moisan, 2010; Wilbraham, 1848; Gibbs, 1899). This is known as the Gibbs-Wilbraham Phenomenon and appears as lines extending from the origin of a 2D Fourier Transform (Figure 1c) (Thomson & Reid, 1927; E Hewitt & R E Hewitt, 1979).

Additional streaks around high intensity regions such as reciprocal lattice peaks are due to scan noise in the HAADF STEM image and are readily distinguished from the cross pattern artifacts. Scan noise causes streaks in a single direction perpendicular to the fast scan axis, originating from the Bragg spots of an atomic resolution image. These artifacts can be reduced by optimizing the microscope's environment, improving the scan electronics, averaging an image series, or with post processing approaches—as discussed by previous investigators (Kirkland & Thomas, 1996; Muller et al., 2006; Jones & Nellist, 2013; Yankovich et al., 2014; Sang & LeBeau, 2014; Braidy et al., 2012).

In this communication, we demonstrate a substantial reduction of the central cross pattern artifact caused by periodic boundaries through implementation of the Periodic Plus Smooth Decomposition (P+S) method. Developed by Moisan (Moisan, 2010) and inspired from previous work (Beaudoin & Beauchemin, 2002; Saito & Remy, 2006), the P+S technique decomposes the original image into a continuous periodic image and a smoothly varying background. The smooth background is obtained by solving Poisson's equation with boundary conditions set by the discontinuities at the image's edges. The periodic image is then obtained by subtracting the smooth varying background and has a Fourier transform that reflects the original image's DFT, but without the cross pattern artifact. Altering only the low frequencies, this approach provides an accurate estimation of the Fourier spectrum in regions were atomic lattice Fourier peaks occur—with contributions from the entire image field. Traditional approaches, such as windowed Fourier Transforms heavily weight information at the center of an image and can miss important structure residing near the edges.

RESULTS

*Traditional Approaches to Periodic Edge Artifact Reduction*

One approach to obtaining continuity at the periodic boundaries is to symmetrize the image. Mirroring an image about the *x,y* directions ensures continuous boundary conditions and greatly reduces the cross pattern artifact in the DFT (Fig. 2b,g). However, this process adds symmetries that do not exist in the original image. As seen in the DFT of the symmetrized image (Figure 2g), additional lattice spots appear that do not reflect the actual specimen. For the analysis of polycrystalline samples, where the radially integrated DFT intensity is of interest, symmetrizing may still be beneficial. Note that because the image size quadruples, there is a computational penalty—albeit manageable.

A more common technique used in periodic artifact reduction is a windowed Fourier transform. The window is typically a smooth and continuous damping function with a central maximum that decays to zero (or nearly zero) at the edges of the image—as shown in Figure 2c,d. This not only removes the edge discontinuities—thus reducing the cross pattern artifacts—but also reduces the background level (side lobes) around each Fourier peak (Figure 2h,i). A multitude of window functions, each named to honor their inventor, have been devised to achieve different advantages (Harris, 1978). Figures 2c,h show the effect of applying a Hann window, a popular general use function, built from a raised cosine function that emphasizes a large central region of the image. As the window function becomes more spatially confined in real space, it degrades the resolution in reciprocal space. For atomic resolution images, this causes the diffraction peaks to become blurred. For closely spaced peaks—like those produced from the NiO [111] twin boundary in Figure 2—the intensity can smear into a single peak. Figure 2d shows a very tight window (flat top Fig. 2i) that causes the pairs of peaks in the DFT (clearly seen in Fig 2f or j) to become difficult or impossible to distinguish.

A more deceptive consequence of using windowing for artifact reduction is the unequal weighting it places on the image. Regions toward the center of the image contribute strongly to the Fourier spectrum and information toward the edges of the field of view becomes negligible. As a result, crystallographic information that does not lie at the center of the image can be easily missed. Figure 3 shows a layered $LaVO_3/SrTiO_3$ material. One of the $LaVO_3$ layers exhibits a super-lattice structure. From the real-space image (Figure 3a), this super-lattice structure is difficult to detect, but it clearly appears as additional peaks in the Fourier transform (Figure 3d). However, this structure would be missed entirely in the windowed transform (Figure 3e)—even when applying the relatively large Hann window.

### *Periodic Plus Smooth Decomposition*

Recently, Moisan (Moisan, 2010) proposed a decomposition method that removes the boundary artifacts (Figure 2j) while preserving all the diffraction peaks from the entire image field of view (Figure 3f). Here, an image ($u$) is decomposed into a sum of a "periodic" component ($p$) and a "smooth" component ($s$), that is, $u = p + s$. Mathematically, the smooth component ($s$) is obtained by solving Poisson's equation $\nabla^2 s = b$, where $b$ is a boundary discontinuity image that captures the intensity gaps across the edges of the original image. In general, Poisson's equation has a unique solution with slow spatial variation (Figure 4c). By subtracting $s$ from the original image, the periodic edge discontinuities are removed and the information at higher frequencies is preserved.

The smooth component ($s$) is computed directly from the "boundary" image. For an M by N image $u(x, y)$, the "boundary" image is defined as $b = b_1 + b_2$, where

$$b_1 = \begin{cases} u(M-1-x, y) - u(x, y) & \text{if } x = 0 \text{ or } x = M-1 \\ 0 & \text{else} \end{cases},$$

$$b_2 = \begin{cases} u(x, N-1-y) - u(x, y) & \text{if } y = 0 \text{ or } y = N-1 \\ 0 & \text{else} \end{cases}.$$

Poisson's equation can be readily solved from the DFT of the "boundary" image, $\hat{b}(k_x, k_y)$, (see Methods):

$$\hat{s}(k_x, k_y) = \frac{\hat{b}(k_x, k_y)}{2\cos\frac{2\pi k_x}{M} + 2\cos\frac{2\pi k_y}{N} - 4} \quad \forall (k_x, k_y) \in \Omega \setminus (0,0),$$

where $\Omega = \{0, \ldots, M-1\} \times \{0, \ldots, N-1\}$ and $\hat{s}(k_x, k_y)$ is the DFT of the smooth component. Additionally, the P+S method imposes the constraint $\hat{s}(0,0) = 0$ so that the mean intensity of the periodic component is equivalent to the original image. The boundary artifact free DFT can be simply computed by $\hat{p}(k_x, k_y) = \hat{u}(k_x, k_y) - \hat{s}(k_x, k_y)$, as illustrated in Figure 2j, 3f, and 4e. Hence, the effective computation cost of performing a P+S Fourier transform is one additional DFT.

As an illustration, Figure 4a shows a STEM image of CdSe nanoparticles. After decomposition, the periodic component (Figure 4b) is continuous across the edges and its Fourier spectrum (Figure 4e) is visually identical to the original image (Figure 4d) with the cross artifacts removed—artifacts that now reside in the smooth component (Figure 4c,f). Since the Fourier transform is a linear operator, the Fourier transform of the smooth component (Figure 4f) illustrates exactly the intensity that has been removed from the image DFT (Figure 4d). The Periodic Plus Smooth Decomposition is also beneficial to high-resolution transmission electron microscopy (HRTEM) (Figure 5). However, HRTEM transfers little information at low frequencies thus discontinuities in the periodic boundaries—and the artifacts produced thereof—are often less prominent than in STEM. The CdSe nanoparticle in Figure 4 originally appeared in H. Zhang et al (Zhang et al., 2011).

METHODS:

In P+S decomposition, the smooth component(s) is computed by solving Poisson's equation $\nabla^2 s = \frac{\partial^2 s}{\partial x^2} + \frac{\partial^2 s}{\partial y^2} = b$. The discrete form of the differential equation is

$$s(x+1, y) - s(x-1, y) + s(x, y+1) - s(x, y-1) - 4s(x, y) = b(x, y).$$

Taking the DFT on both sides and invoking the shift theorem, one obtains

$$\left(e^{-\frac{2\pi i k_x}{M}} + e^{\frac{2\pi i k_x}{M}} + e^{-\frac{2\pi i k_y}{N}} + e^{\frac{2\pi i k_y}{N}} - 4\right) \hat{s}(k_x, k_y) = \hat{b}(k_x, k_y),$$

where $\hat{s}(k_x, k_y)$ and $\hat{b}(k_x, k_y)$ are the DFT of $s(x, y)$ and $b(x, y)$, respectively.

Thus, for all $(k_x, k_y) \in \{0, \ldots, M-1\} \times \{0, \ldots, N-1\} \setminus (0,0)$,

$$\hat{s}(k_x, k_y) = \frac{\hat{b}(k_x, k_y)}{2\cos\frac{2\pi k_x}{M} + 2\cos\frac{2\pi k_y}{N} - 4}$$

The P+S decomposition and the other methods in this manuscript are implemented in Python and available in the Supplemental Material.

## Conclusion

Here, we demonstrated that Periodic Plus Smooth Decomposition is a promising technique for Fourier analysis of atomic resolution images by effectively removing the periodic boundary artifact implicit to a standard DFT. The traditional approach requires application of a damping window to the image that degrades resolution in Fourier space and, more critically, can omit useful information toward the edges of the field of view. Periodic Plus Smooth Decomposition utilizes Poisson's equation with boundary conditions set by the discontinuities at the edges of the image to find a smooth, slowly varying background. By separating this smooth background from the original image, Periodic Plus Smooth Decomposition provides a Fourier transform of the entire image, at full spectral resolution, and free from the cross pattern artifact caused by boundary discontinuities. This technique provides a valuable tool to aid routine Fourier analysis of high-resolution electron micrographs.

Acknowledgements

This work was supported in part by the Cornell Center for Materials Research with funding from the NSF MRSEC program (DMR-1120296). H.L.X. is supported by the Center for Functional Nanomaterials, Brookhaven National Laboratory, which is supported by the U.S. DOE, Office of Basic Energy Sciences, under Contract No. DE-AC02-98CH10886. Y.J. is supported by DOE grant DE-FG02-11ER16210. We thank T. Higuchi, H. Y. Hwang at Stanford University for providing the sample used in Fig. 3.

# FIGURE CAPTIONS

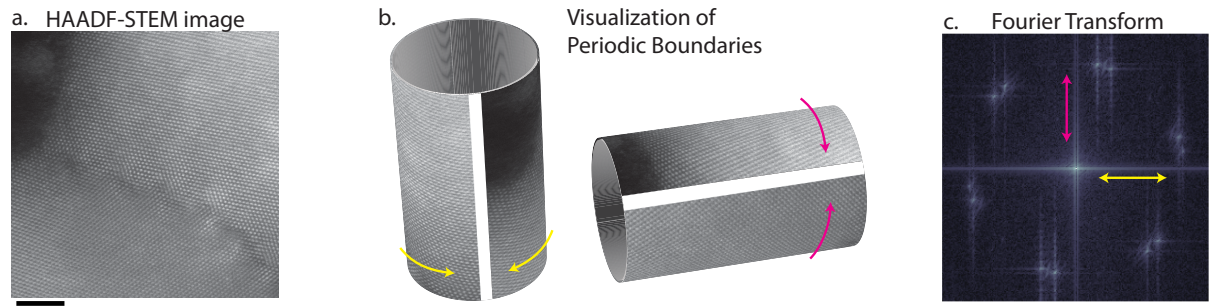

Figure 1 | Schematic illustration of the periodic boundary condition imposed by the discrete Fourier transform of an image (a) along the vertical and horizontal directions (b). Discontinuities at the boundaries cause cross pattern artifacts in the Fourier transform (c). Scale bar 2 nm.

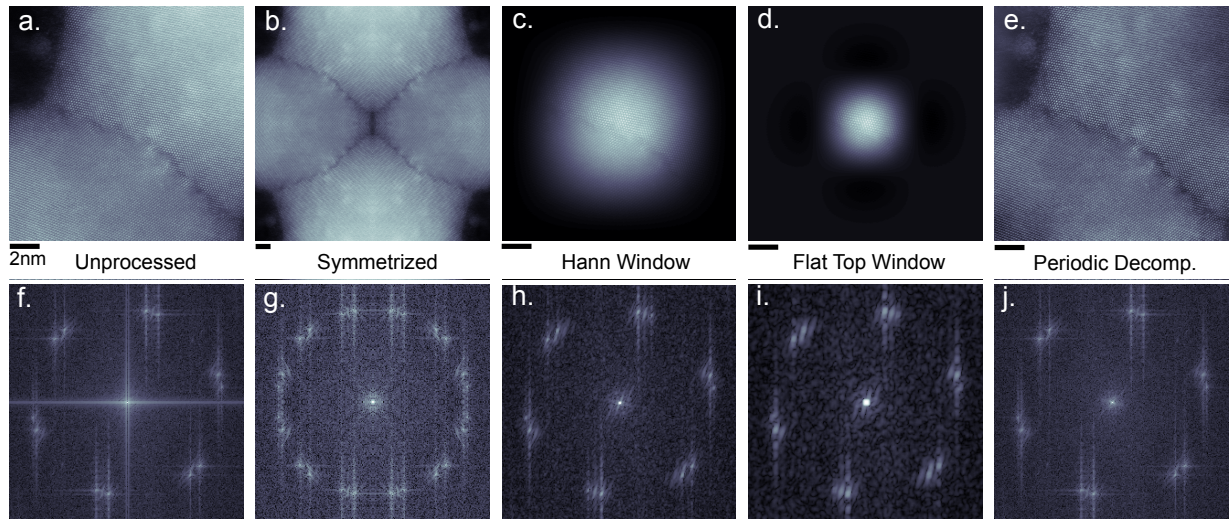

Figure 2 | Comparison of approaches to reduce periodic boundary artifacts in the discrete Fourier transforms (DFT) of a HAADF-STEM image. A twin boundary in the NiO [111] gives rise to six pairs of Bragg peaks. The DFT of an unprocessed image (a.) contains vertical and horizontal boundary artifacts (f.). Symmetrizing the image (b.) removes these artifacts, but additional Bragg spots appear (g.). A Hann window (c.) is effective at removing the boundary artifacts (h.) but emphasizes only the central region of the image (c.). If the window is too tight, like that of the Flat Top (d.) the Bragg peak pairs blur together and become indistinguishable (i.). The boundary artifacts can be effectively removed using Periodic Plus Smooth Decomposition (j.) while preserving the image's entire field of view (e.) and resolution in reciprocal space (j.). Scale bars are 2nm and the {220}; Fourier peaks reflect a 0.68 Å$^{-1}$ spacing; DFT images are log-absolute intensities.

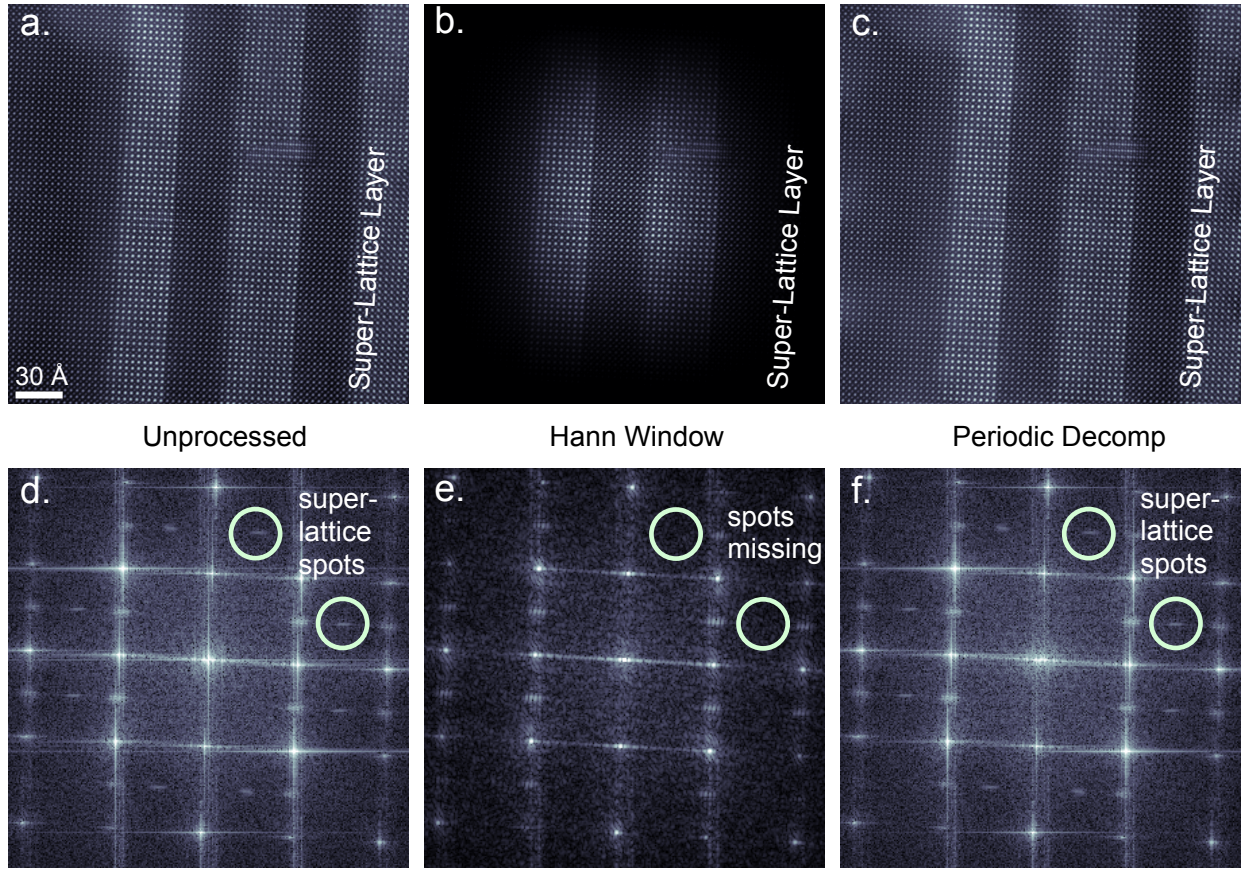

Figure 3 | Layered LaVO$_3$/SrTiO$_3$ material with one LaVO$_3$ layer containing a super-lattice (a.) that appears as additional peaks in the Fourier transform (d.). A windowed Fourier transform heavily weights contributions at the center of the image (b.) and consequently misses the super-lattice peaks that originate from the edge regions of the image (e.). The Periodic Plus Smooth Decomposition utilizes the entire field of view (c.), thus, preserving the atomic superlattice structure in the DFT (f.). DFT images (d-f.) are log intensity and cropped—the superlattice spots occur at 0.23 Å$^{-1}$. Sample was grown by T. Higuchi, H. Y. Hwang, Stanford University.

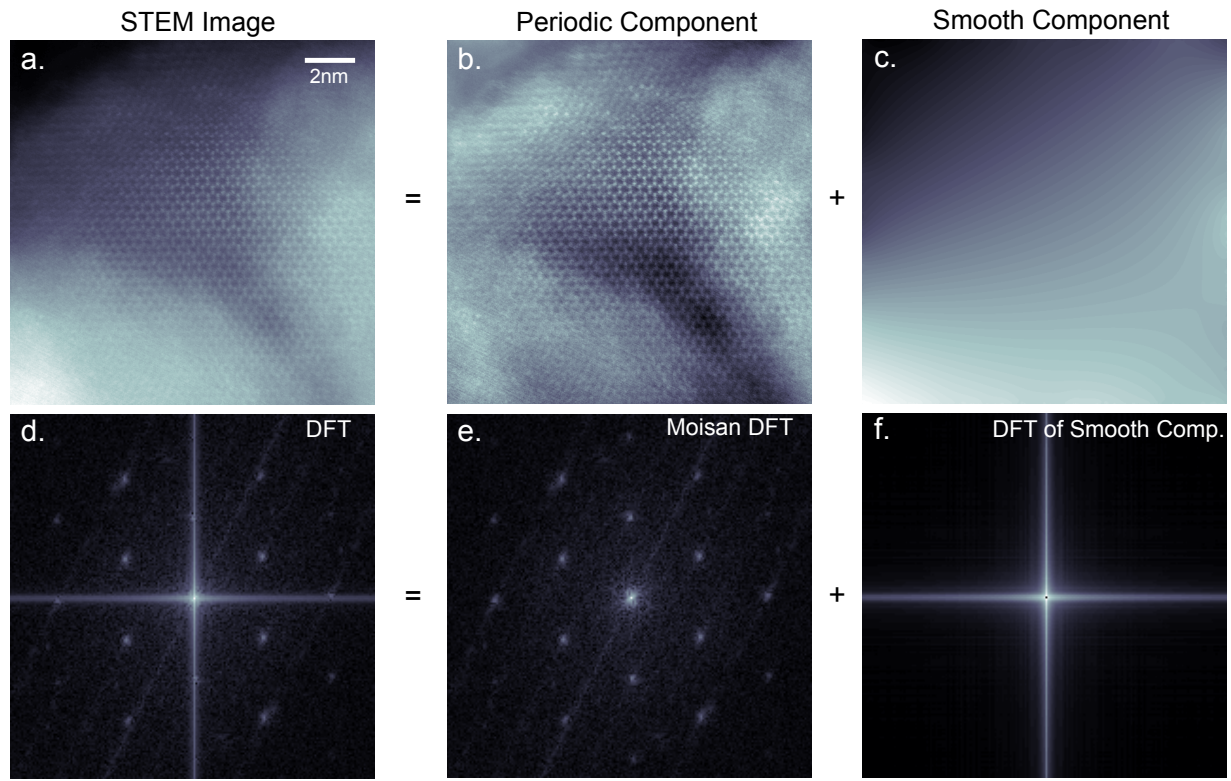

Figure 4 | Periodic Plus Smooth Decomposition of a CdSe [111] nanoparticle resting between two other particles. The STEM image contains strong discontinuities in the periodic boundaries manifesting as cross pattern artifacts in the DFT (d.). The P+S decomposition creates continuity in the periodic boundaries (b.) by removing a smooth background component (c.) from the image. This removes the boundary artifacts in the DFT (e.). The DFT of background component (f.) shows exactly what has been removed from the DFT of the original image (d.) to create the artifact free DFT (e.). DFT images are log-absolute intensities; scan direction is rotated 26.5° clockwise from horizontal.

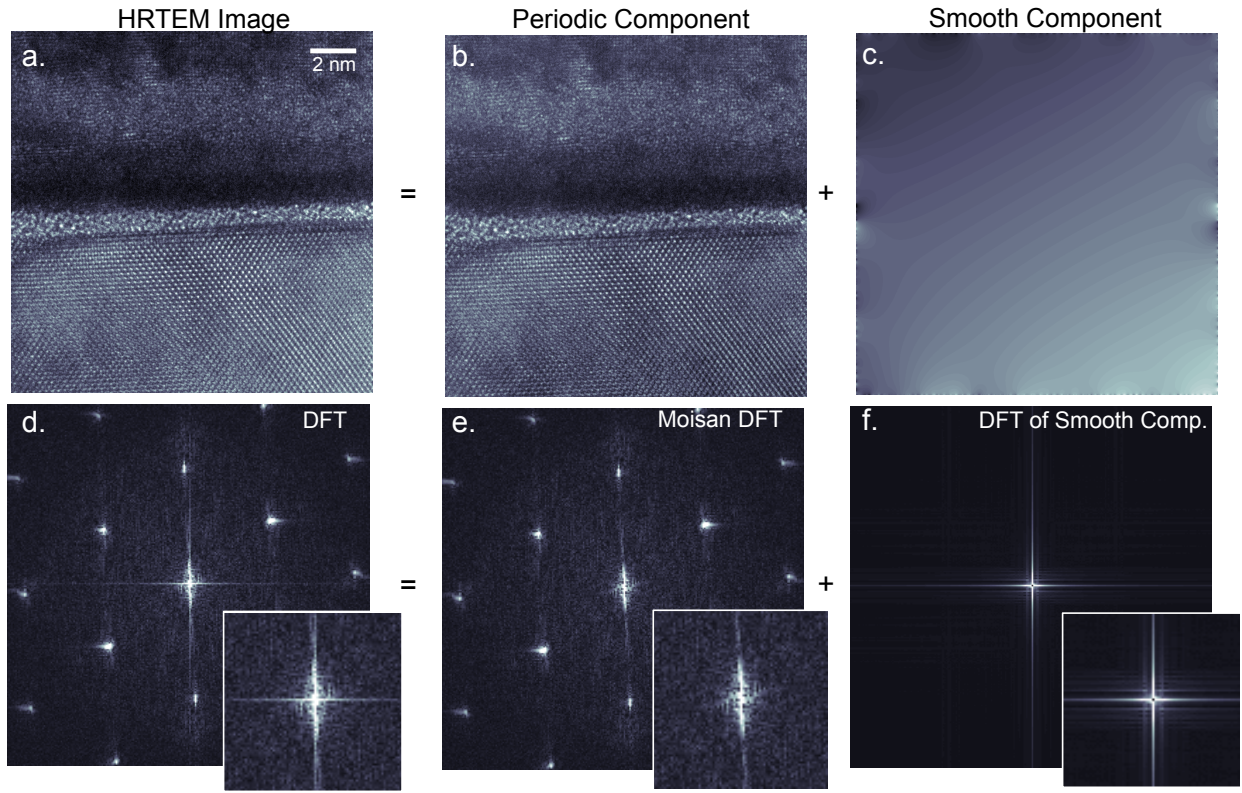

Figure 5 | Periodic Plus Smooth Decomposition of a semiconductor high-k oxide on a silicon substrate (viewed along the Si [110] direction) imaged by HRTEM. The HRTEM image transfers little information at low frequencies; however, small discontinuities in the periodic boundaries at the image edges are still present and manifest as a cross pattern artifact in the DFT (d.). The P+S decomposition removes the boundary artifacts in the DFT (e.). The DFT of background component (f.) shows the artifact removed from the original DFT (d.) DFT images are log-absolute intensities; DFT images have been cropped and contain insets of lowest frequencies where artifacts are most noticeable.

## FOURIER TRANSFORM OF SYMMETRIZED IMAGE IN PYTHON (2D)

```python
from pylab import *
#This function returns the Symmetrized Image FFT
def symfft2(im):
    symim = zeros( [2*size(im,0), 2*size(im,1)] )
    symim[0:size(im,0),0:size(im,1)] = im
    symim[0:size(im,0),size(im,1):2*size(im,1)] = fliplr(im)
    symim[size(im,0):2*size(im,0),0:size(im,1)] = flipud(im)
    symim[size(im,0):2*size(im,0),size(im,1):2*size(im,1)] = flipud(fliplr(im))
    return( fft2(symim) )
```

## WINDOWED FOURIER TRANSFORM IN PYTHON (2D)

```python
from pylab import *
#This function returns the Hann Windowed Fourier Transform
def hannfft(im):
    window = outer( hanning(size(im,0)), hanning(size(im,1))
    winim = window * im
    return( fft2(winim) )
```

## PERIODIC PLUS SMOOTH DECOMPOSITION IN PYTHON (2D)

```python
from pylab import *
#This function returns the reciprocal space P and S components
def mfft2(im):
    [rows,cols] = shape(im)
```

```python
    #Compute boundary conditions
    s = zeros( shape(im) )
    s[0,0:]      = im[0,0:] - im[rows-1,0:]
    s[rows-1,0:] = -s[0,0:]
    s[0:,0]      =  s[0:,0] + im[0:,0] - im[:,cols-1]
    s[0:,cols-1] =  s[0:,cols-1] - im[0:,0] + im[:,cols-1]
    #Create grid for computing Poisson solution
    [cx, cy] = meshgrid(2*pi*arange(0,cols)/cols,
2*pi*arange(0,rows)/rows)

    #Generate smooth component from Poisson Eq with boundary condition
    D = (2*(2 - cos(cx) - cos(cy)))
    D[0,0] = inf    # Enforce 0 mean & handle div by zero
    S = fft2(s)/D

    P = fft2(im) - S # FFT of periodic component
    return(P,S)
```